\documentclass[aps,showpacs,floatfix,footinbib,preprintnumbers]{revtex4}
\pdfoutput=1
\usepackage{color,amsmath,amssymb,graphicx,latexsym,subfigure}
\usepackage{txfonts}

\usepackage{hyperref}

\newcommand{\sv}{\ensuremath{\langle\sigma v_{\text{rel}}\rangle}}
\newcommand{\sigsip}{\ensuremath{\sigma_{\chi p}^{\rm{SI}}}}

\definecolor{DeepPink}{rgb}{1.0, 0.08, 0.58}

\begin{document}

\title{
A model explaining neutrino masses and the DAMPE cosmic ray electron excess
}

% NCTS-PH/17xx \;\;

\preprint{NCTS-PH/1727, CP3-Origins-2017-056 DNRF90\\}

\author{Yi-Zhong Fan$^{a,b}$\footnote{E-mail: \texttt{yzfan@pmo.ac.cn}}}
\author{Wei-Chih Huang$^c$\footnote{E-mail: \texttt{huang@cp3.sdu.dk}}}
\author{Martin Spinrath$^d$\footnote{E-mail: \texttt{martin.spinrath@cts.nthu.edu.tw}}}
\author{Yue-Lin Sming Tsai$^{e}$\footnote{E-mail: \texttt{smingtsai@gate.sinica.edu.tw}}}
\author{Qiang Yuan$^{a,b}$\footnote{E-mail: \texttt{yuanq@pmo.ac.cn}}}

\affiliation{
$^{a}$Key Laboratory of Dark Matter and Space Astronomy, Purple Mountain Observatory, Chinese Academy of Sciences, Nanjing 210008, China\\
$^{b}$School of Astronomy and Space Science, University of Science and Technology of China, Hefei, Anhui 230026, China\\
$^{c}$CP$^3$ Origins, University of Southern Denmark, Campusvej 55, DK-5230 Odense M, Denmark\\
$^{d}$Physics Division, National Center for Theoretical Sciences, Hsinchu 30013, Taiwan\\
$^{e}$Institute of Physics, Academia Sinica, Nangang, Taipei 11529, Taiwan}

\begin{abstract}
We propose a flavored $U(1)_{e\mu}$ neutrino mass and dark matter~(DM) model to
explain the recent DArk Matter Particle Explorer (DAMPE) data, which feature an excess 
on the cosmic ray electron plus positron flux around 1.4 TeV.
Only the first two lepton generations of the Standard Model are charged 
under the new $U(1)_{e\mu}$ gauge symmetry. A vector-like fermion $\psi$,
which is our DM candidate, annihilates into $e^{\pm}$ and $\mu^{\pm}$ via
the new gauge boson $Z'$ exchange and accounts for the DAMPE excess.
We have found that the data favors a $\psi$ mass around 1.5~TeV and
a $Z'$ mass around 2.6~TeV, which can potentially be probed by 
the next generation lepton colliders and DM direct detection experiments. 
\end{abstract}

\maketitle

\setcounter{footnote}{0}
\renewcommand{\thefootnote}{\arabic{footnote}}

\section{Introduction}
%{\it Introduction ---}
The newly released data from the DArk Matter Particle Explorer~(DAMPE~\cite{TheDAMPE:2017dtc}) exhibits an intriguing excess of the cosmic ray
electron plus positron (hereafter CRE) flux at energies around 1.4 
TeV~\cite{Ambrosi:2017wek}. We provide here a dark matter (DM) explanation 
based on a simple flavored $U(1)$ extension of the standard model (SM). 
This kind of extension is known for quite a 
while~\cite{He:1990pn, Foot:1990mn, He:1991qd}. Well-studied 
scenarios are those involving the second and third generation, 
$U(1)_{\mu\tau}$~(denoted as $L_\mu - L_{\tau}$ in the literature), which are 
partially motivated by the large mixing angle inferred from atmospheric 
neutrino oscillations~\cite{Heeck:2011wj,Baek:2015mna,Biswas:2016yan}. 
Such models are recently used to explain anomalies in 
Higgs and quark flavor physics (see, e.g.~\cite{Altmannshofer:2014cfa, 
Heeck:2014qea, Crivellin:2015mga, Crivellin:2015lwa}). This class
of models was also discussed in the context of the PAMELA, ATIC and FERMI
results \cite{Bi:2009uj}.

In this work, we focus on another variant, $U(1)_{e\mu}$, under which only 
the first two generation leptons are charged. This choice is inspired by 
DAMPE CRE data as we are trying to establish the connection between the 
DM explanation for the CRE excess and the neutrino mass generation mechanism. 
In this framework, the DM candidate is a  vector-like fermion $\psi$ whose stability is guaranteed by an accidental 
$U(1)$ symmetry. The DM annihilation into $e^\pm$ and 
$\mu^\pm$~(also $\nu_e$ and $\nu_\mu$) can account well for the 
DAMPE excess. Since the generated electrons and positrons 
lose energies quickly on the way to the Earth, the CREs detected by 
DAMPE must come from regions close to the solar neighborhood. 
As a result, we assume that there exists a nearby DM subhalo, which is
also predicted by the structure formation of the cold DM scenario
(e.g.~\cite{Gao:2004au,Diemand:2005vz}).

\section{The model}

% {\it The model ---}
Our model is a rather minimal extension of the SM. We add one additional
anomaly-free $U(1)_{e\mu}$ gauge group, two additional scalars, $\phi_1$ 
and $\phi_2$, whose vacuum expectation values (vevs) break the new $U(1)_{e\mu}$ spontaneously, 
three right-handed neutrinos, and a vector-like fermion $\psi$ as a DM 
candidate. Only the lepton doublets, right-handed leptons and neutrinos 
of the first two generations are charged under $U(1)_{e\mu}$ as summarized in 
Table~\ref{tab:Model}. The fermion $\psi$ is stable since the 
Lagrangian carries an additional accidental $U(1)$ symmetry which can 
be interpreted as $\psi$-number.

\begin{table}
\centering
\caption{\label{tab:Model}
Charge assignments of the fields under the new $U(1)_{e\mu}$ gauge group
which is broken by the vevs of the scalar fields $\phi_1$ and $\phi_2$.
The fermion $\psi$ is our DM candidate. These three new fields do not carry any SM 
quantum numbers and all the SM fields not shown are neutral under $U(1)_{e\mu}$.}

\begin{tabular}{lcccccccccc}
\hline
Field & $L_e$ & $L_\mu$ & $e_R$ & $\mu_R$ & $N_1$ & $N_2$ & $N_3$ & $\psi$ & $\phi_1$ & $\phi_2$ \\
\hline 
$U(1)_{e\mu}$ charge & $1$ & $-1$ & $1$ & $-1$ & $1$ & $-1$ & $0$ & $q_\psi$ & $1$ & $2$  \\
\hline
\end{tabular}
\end{table}

In this model, the $U(1)_{e\mu}$ symmetry demands both the charged lepton and the neutrino Yukawa
couplings to be diagonal in the flavor basis. 
On the other hand,  when the scalars
receive a vev the resulting right-handed neutrino mass matrix is an unconstrained 
symmetric matrix: 
\begin{align}
M_R = \frac{1}{2}
\begin{pmatrix}
  y_{11} \langle \phi_2^* \rangle & M_{12} & y_{13} \langle \phi_1^* \rangle     \\
  M_{12} &  y_{22} \langle \phi_2 \rangle & y_{23}\langle \phi_1 \rangle \\
  y_{13} \langle \phi_1^* \rangle  & y_{23} \langle \phi_1 \rangle  & M_3
\end{pmatrix} \, ,
\end{align}
where $y_{ij}$ are Yukawa couplings of the right-handed neutrinos with 
the scalar singlets $\phi_1$ and $\phi_2$, and $M_{12}$ and $M_3$ are
mass parameters.
With such structures we can reproduce the neutrino masses and mixing angles
via the Type-I seesaw mechanism.

The scalar potential in the unbroken phase reads
\begin{align}
 V_s &= - \mu_H^2 \, |H|^2 + \lambda_H \, |H|^4 - \mu_{\phi_i}^2 \, |\phi_i|^2 + \lambda_i \, |\phi_i|^4 \nonumber\\
 & + \lambda_{12} \, |\phi_1|^2 |\phi_2|^2 + \kappa_i \, |\phi_i|^2 |H|^2 \;,
\end{align}
where $H$ is the usual SM Higgs doublet, and we have $\mu_H^2 > 0$ and 
$\mu_{\phi_i}^2 > 0$ for $i = 1,2$. After electroweak and $U(1)_{e\mu}$
symmetry breaking, $\langle H^0 \rangle  
= v_H/\sqrt{2}$ and $\langle \phi_i \rangle  = v_{\phi_i}/\sqrt{2}$, 
there exist three physical CP even Higgs bosons $h$ and $\eta_i$ with 
masses $m_h$ and $m_{\eta_i}$, and one CP odd Higgs boson $\zeta$ with 
a mass $m_\zeta$. For simplicity, we assume here that the $\kappa_i$ are negligibly 
small so that $h$ is identified with the SM Higgs boson. 
The $\kappa_i$ terms could be probed with future Higgs precision data.
A careful and detailed study is, however, beyond the scope of this work.

The mass of the new gauge boson is $m_{Z'}^2 = g_{e\mu}^2 (v_1^2 + 4 v_2^2)$ 
on tree level, where $g_{e\mu}$ is the $U(1)_{e\mu}$ gauge coupling.
Since the $\phi_i$ do not carry any SM quantum numbers, the masses of the SM 
gauge bosons are not affected by $\langle \phi_i \rangle$ on tree level.

The relevant Lagrangian for the DM annihilation into SM fermions $f$ is
\begin{align}
  \mathcal{L} \supset & - \frac{1}{2} m^2_{Z'} Z'^\mu Z'_\mu - m_\psi \bar{\psi} \psi + \text{i}\, q_\psi g_{e\mu} \bar{\psi}\gamma^\mu{\psi} Z'_{\mu} \nonumber \\
   & - \sum_{f=e, \mu} \left( m_f \bar{f} f 
  - \text{i}\, q_f g_{e\mu} \bar{f}\gamma^\mu{f} Z'_{\mu} \right) \nonumber \\
& +    \sum_{f=\nu_e, \nu_\mu}  
 \text{i}\,  q_f g_{e\mu}   \bar{f}\gamma^\mu  \left( \frac{1-\gamma_5}{2}\right) {f} Z'_{\mu}
   \, ,
  \end{align}
where $q_f$ labels the $U(1)_{e\mu}$ charge of the field $f$, 
c.f.~Table~\ref{tab:Model}.  
The SM fermion masses are neglected due to $m_f \ll m_\psi$ for our regions of interest.
We further assume that 
the extra scalars, $\eta_i$ and $\zeta$, and the right-handed neutrinos 
are all heavier than $\psi$ and $Z'$.

The DM annihilation cross-section into a SM fermion pair $\bar{f}f$,
$\sigma(\bar{\psi} \psi \to Z' \to \bar{f} f)$, multiplied by the DM
relative velocity $v_{\text{rel}}$, is
 \begin{align}
 \sigma v_{\text{rel}} = c_f \frac{q^2_\psi q_f^2 \, g_{e\mu}^4  \left( s + 2 \, m^2_\psi\right)}{
 6 \pi \left[   \left(  s - m^2_{Z'} \right)^2 + m^2_{Z'}\Gamma^2_{Z'} \right] 
 } \, ,
 \end{align}
where $c_f$=1~(1/2) for $e$ and $\mu$~($\nu_e$ and $\nu_\mu$),
and 
$s=16\,m^2_\psi/(4-v^2_{\text{rel}})$ is the square of the center-of-mass 
energy. Note that $\sigma v_{\text{rel}}$ is dominated by the $s$-wave 
component as $v_{\text{rel}} \to 0$.
The total $Z'$ decay width into $\bar{f}f$ and $\bar{\psi}\psi$ reads
\begin{align}
 \Gamma_{Z'} & = \sum_{f=e, \mu, \nu_e, \nu_\mu} c_f
 \frac{q_f^2 g_{e\mu}^2 m_{Z'}}{12 \, \pi} \nonumber \\
 &+ \Theta(m_{Z'} - 2 \, m_{\psi}) \frac{q_\psi^2 g_{e\mu}^2 \sqrt{m^2_{Z'} - 4 \, m^2_\psi } \left( m^2_{Z'} + 2 \, m^2_\psi  \right) }{12 \, \pi \, m^2_{Z'}} 
 \, .
\end{align}

\section{Parameter space}
%{\it Parameter space ---}
We first study the CRE background, i.e., the CRE not from DM annihilations. 
The background component (from astrophysical 
sources such as supernova remnants and/or pulsars) is assumed to have 
a double-broken power-law form as
\begin{equation}
\Phi_{\rm bkg}=\Phi_0\,E^{-\gamma}\left[1+\left(\frac{E_{\rm br,1}}{E}
\right)^{\delta}\right]^{\Delta\gamma_1/\delta}\,
\left[1+\left(\frac{E}{E_{\rm br,2}}\right)^{\delta}\right]
^{\Delta\gamma_2/\delta},
\end{equation}
with the first break at $E_{\rm br,1}\sim50$ GeV and the second one at 
$E_{\rm br,2}\sim900$ GeV according to the Fermi-LAT~\cite{Abdollahi:2017nat}
and DAMPE observations~\cite{Ambrosi:2017wek}. During the analysis, we fix $E_{\rm br,1}$ to 
50~GeV, and the sharpness parameter $\delta$ to 10. 
The fit to the DAMPE data with the $e^\pm$ energy between 25~GeV and 4.6~TeV without taking into account
the peak~(excess) 
leads to $\Phi_0=247.2$ GeV$^{-1}$~m$^{-2}$~s$^{-1}$~sr$^{-1}$, $\gamma=3.092$, 
$\Delta\gamma_1=0.096$, $\Delta\gamma_2=-0.968$, and $E_{\rm br,2}=885.4$~GeV.

Next, we include the contribution from a nearby DM subhalo in addition to the background 
and fit again to the data. The density distribution inside the subhalo is
assumed to be a Navarro-Frenk-White profile~\cite{Navarro:1996gj}, with
a truncation at the tidal radius $r_t$~\cite{Springel:2008cc}. For the
determination of the density profile of the subhalo, we refer to
Ref.~\cite{Yuan:2017ysv}. As for the propagation of electrons and 
positrons in the Milky Way, we adopt the Green's function approach presented 
in Ref.~\cite{Atoyan:1995}.

\begin{figure}
\centering
\includegraphics[width=0.45\textwidth]{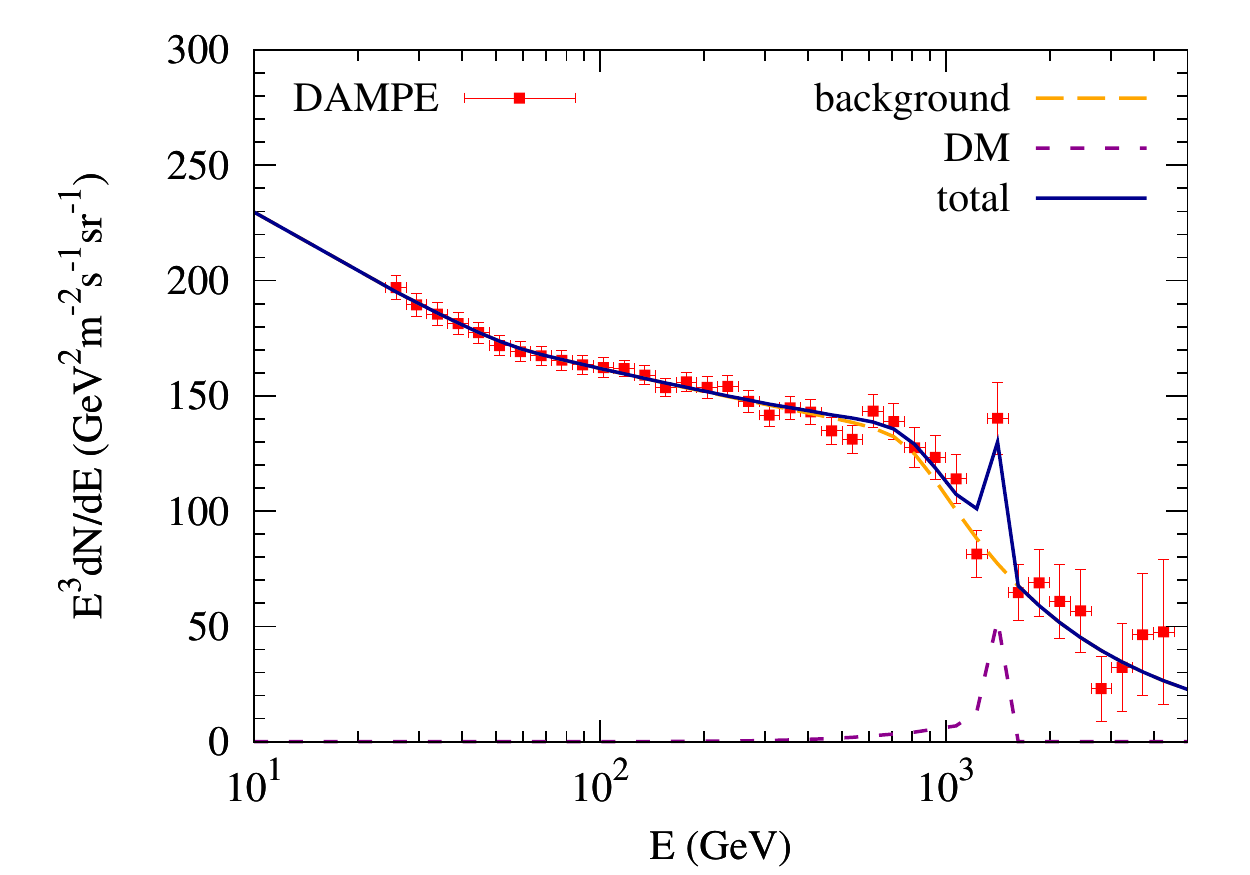}
\caption{Illustration of the fit to the total DAMPE CRE flux with
the background and the DM contribution.
\label{fig:elec_fit_illu}}
\end{figure}

The background parameters $E_{\rm br,2}$ and $\Delta\gamma_2$ are 
correlated to the DM component, and thus are being varied in the fit. Other 
parameters are fixed to the best-fit values obtained in the aforementioned 
background-only fit. Fig.~\ref{fig:elec_fit_illu} shows
 the model prediction of the CRE flux for $m_{\psi}=1.54$ TeV, 
$\sv=6.82\times10^{-25}$ cm$^3$~s$^{-1}$, and the DM subhalo with a mass of
$M_{\rm sub}=1.25\times10^6$ M$_{\odot}$ at a distance of $d=0.1$ kpc 
from the Earth.

There are four relevant DM parameters in this model: $m_\psi$, 
$m_{Z'}$, $g_{e\mu}$, and $q_\psi$.
To ensure the DM model withstands various experimental bounds and to 
explore favored regions of the parameter space, we consider the 
constraints from (i) the relic density, 
(ii) the cosmic microwave background (CMB), 
(iii) the LEP measurements on the cross-sections of 
the leptonic final states, (iv) DM direct detection,
and (v) the DAMPE data.
Note that the recent measurements of the CRE flux
by the Calorimetric Electron Telescope (CALET) up to 3 TeV~\cite{Adriani:2017efm}
are not considered here, because of the relatively large statistical and systematic 
uncertainties. For the relic density, we use the Planck result: 
$\Omega_\psi h^2=0.1199\pm 0.0027$~\cite{Ade:2015xua} plus $10\%$ 
theoretical uncertainties, which are commonly included to 
take into account the discrepancies among the different Boltzmann equation solvers and entropy tables.

The constraints on the DM annihilation rate from the PLANCK 
\texttt{TT,TE,EE+lowP} power spectra (Table 6 of Ref.~\cite{Ade:2015xua}) 
are employed. Moreover, the LEP measurements on the cross-section of 
$e^+ e^- \to \ell^+ \ell^-$ can be translated into constraints on the 
new physics scale in the context of the effective four-fermion 
interactions~\cite{LEP:2003aa}
\begin{align}
\mathcal{L}_{\text{eff}}= \frac{4\pi}{ \left( 1 + \delta \right) \Lambda^2} \sum_{i,j=L,R} \eta_{i,j} \bar{e}_i \gamma_\mu e_i \bar{f}_j \gamma^\mu f_j \, ,
\end{align}
where $\delta=0~(1)$ for $f\neq e~(f=e)$, and $\eta_{ij}=1~(-1)$ 
corresponds to constructive~(destructive) interference between the SM 
and new physics processes. For $e^+ e^- \to e^+ e^-$~($e^+ e^- 
\to \mu^+ \mu^-$), one has $\Lambda=18$~(21.7)~TeV, which implies
$m_{Z'}/g_{e\mu}\gtrsim 7.2~(6.1)$~TeV.

 Even if DM couples only to leptons at tree level, spin-independent 
DM-proton interactions can still be loop-induced and probed as discussed in, for instance, Refs.~\cite{Kopp:2009et,Huang:2013apa,DEramo:2017zqw}.
A recent updated analysis based on 
a leptophilic dark sector in Ref.~\cite{DEramo:2017zqw} attains the constraints from direct detection on the DM and mediator mass for different types of DM-lepton interactions, as displayed in Fig.~2 therein.
To apply the results to our model,
we take the direct detection constraints in Ref.~\cite{DEramo:2017zqw}
for the vector-type interaction~(solid blue line in their upper-right 
panel of Fig.~2) and then rescale it with our coupling constants. 
To realize our $U(1)_{e\mu}$ model, only the vector couplings for $e$ and $\mu$ are nonzero, $g_{Ve}$, $g_{V\mu} \neq 0$
in the notation of Ref.~\cite{DEramo:2017zqw}.
Furthermore, the direct detection limit given in Ref.~\cite{DEramo:2017zqw} is based on 
the LUX \texttt{WS2014-16} run~\cite{Akerib:2016vxi}  which is slightly less stringent than
that from the latest PandaX-II data~\cite{Cui:2017nnn}.
As a consequence, with the new data the lower bound on the mediator mass will improve  by a factor of $[\sigsip(\rm{LUX})/\sigsip(\rm{PandaX})]^{1/4}$,
given a DM mass. 
With these rescalings taken into account,
the derived bound for $m_{Z'} \sim \mathcal{O}(\text{TeV})$ is
\begin{align}
g^2_{e \mu} q_\psi \left( \frac{1170 \, \text{GeV}}{m_{Z'}} \right)^2 \lesssim 1 \, ,
\end{align} 
where we set $q_{e,\mu}=1$. The XENON1T~\cite{Aprile:2017iyp} 
data yield a similar limit.

The DM particle mass $m_\psi$ in the analysis ranges from $0.5$ to 
$5.0$~TeV with the $Z'$ mass in the range
$m_\psi < m_{Z^\prime} 
< 2 m_\psi$, making the current $\sv$ larger than it was at the time of DM freeze-out, 
although the resonance enhancement needs not to be enormous.
The DM charge $q_\psi$ is  varied between $0.5$ and $5$. 
We conducted a random scan and a
Nest-Sampling scan of the parameter space. 
After identifying the high probability region by checking the result of the random scan, 
we utilized \texttt{MultiNest}~\cite{Feroz:2008xx} in the Nest-Sampling scan 
to optimize the coverage of sampling. 
The two scans ($\sim10^8$ points) are then combined with a profile likelihood method.

\begin{figure}
\centering
\includegraphics[width=0.4\textwidth]{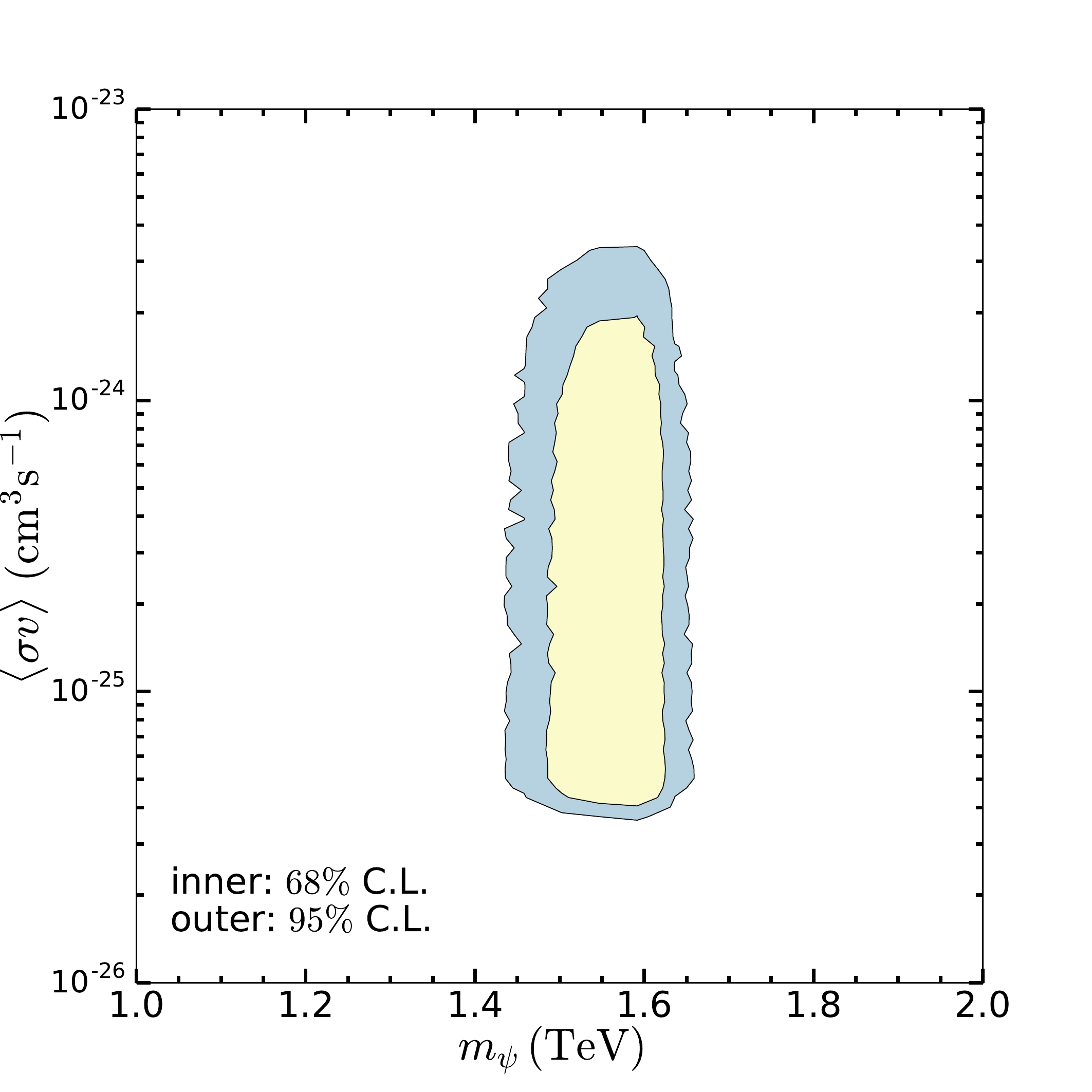}
\caption{The $68\%$ (inner) and $95\%$ (outer) contours for $m_\psi$ versus $\sv$. 
The total likelihoods included here are the relic density, 
the CMB constraints on DM annihilation into charged
leptons, the LEP $Z'$ constraints, the DM direct detection constraints and 
the DAMPE CRE measurement.
\label{fig:mx_sv}}
\end{figure}

\begin{figure}
\centering
\includegraphics[width=0.4\textwidth]{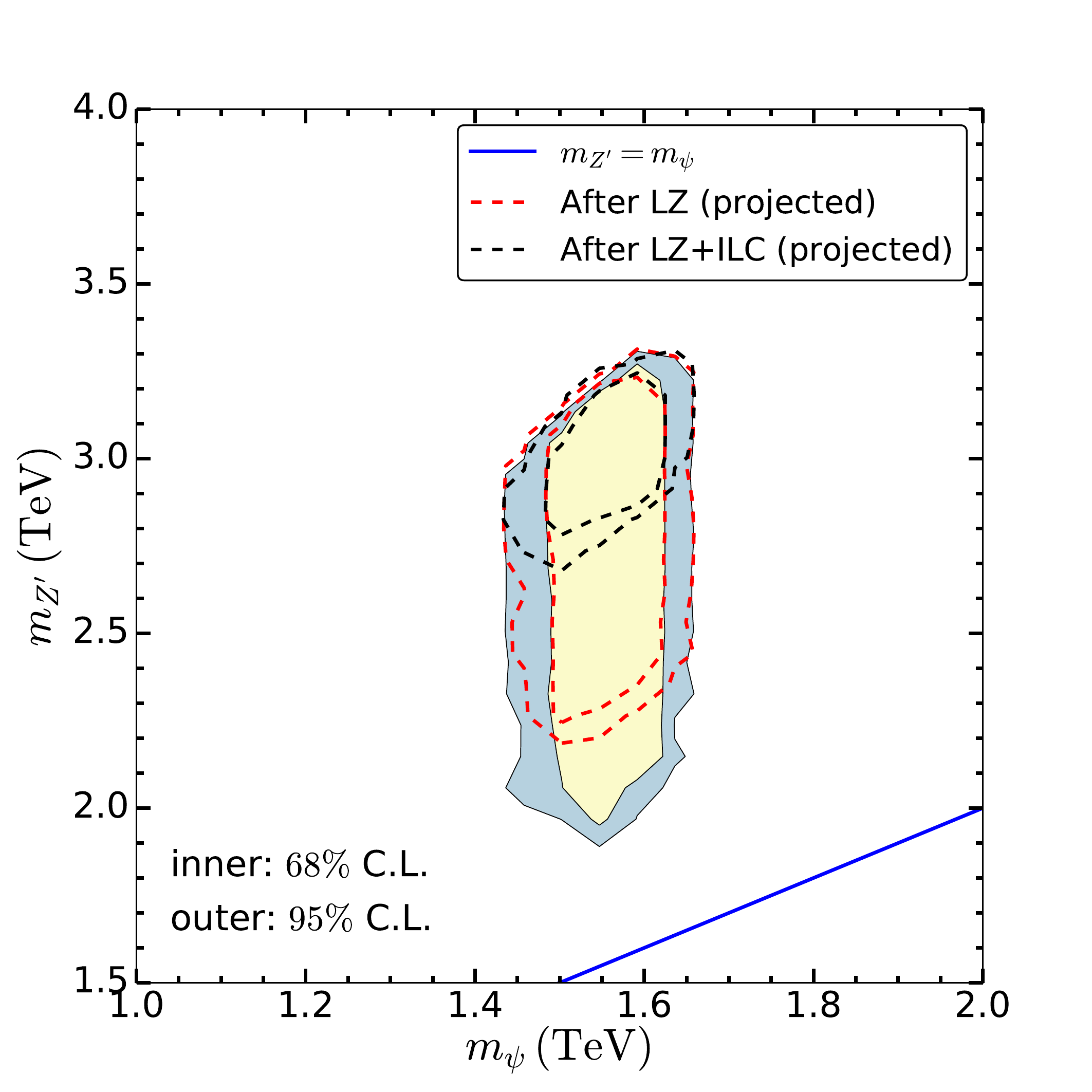}
\caption{The $68\%$ (inner) and $95\%$ (outer) contours for $m_\psi$ versus $m_{Z'}$. 
The filled contours are based on the present constraints as shown in Fig.~\ref{fig:mx_sv}. 
However, for a future prospect, including the projected constraints from LZ (red dashed contours) and 
LZ+ILC (black dashed contours) are presented as well.  
\label{fig:combined}}
\end{figure}

In Fig.~\ref{fig:mx_sv} and Fig.~\ref{fig:combined}, we present the $68\%$ (inner) and $95\%$ (outer) 
profile likelihood contours on the plane of $m_\psi - \sv$ and $m_\psi - m_{Z'}$, respectively.
 The preferred DM mass region is between 1.4~TeV and 
1.7~TeV with a $Z'$ mass between 1.9~TeV and 3.2~TeV and $g_{e\mu}$ between
$0.014$ and $0.38$ at the $95\%$ CL. We find no preferred region for
$q_\psi$ over the scan range $[0.5, \, 5]$.

Together with the coupling limits from PLANCK (relic density and CMB), 
 the DM annihilation cross-section is confined within
$[3\times10^{-26}, \, 3\times10^{-24}]$~cm$^3$~s$^{-1}$ as shown in Fig.~\ref{fig:mx_sv}.
The annihilation  cross-section is inversely proportional to the mass of the subhalo, which 
is restricted inside the range of  $[2.5\times10^5, \, 6\times10^7]$ M$_{\odot}$,
assuming a distance of $d=0.1$ kpc.
For different values of $d$, the required subhalo mass scales 
approximately as $d^2$~\cite{Yuan:2017ysv}.

\section{Other constraints and prospects}
%{\it Other constraints and prospects ---}
We further consider bounds from DM indirect detection, and also comment on
the model's detectability at future DM direct detection and lepton colliders.

\begin{itemize}

\item Fermi-LAT $\gamma$-ray data

We have checked that the inverse
Compton emission from the diffuse electrons and positrons for the presumed  
subhalo is negligibly small. 
On the other hand, we also study the $\gamma$-ray emission produced via the internal 
bremsstrahlung process by the charged fermions which come from the $Z'$ 
decays. The process is known as the final state radiation 
(FSR; \cite{Bergstrom:2004cy}).  The FSR $\gamma$-rays from the
subhalo are essentially extended over a considerable patch of the sky.
The expected numbers of photons from the DM annihilation 
within the subhalo for $E_{\gamma}>100$ GeV are estimated to be 
 0.7, 2.0, 5.9, 13.6, 25.3, 34.1, 34.4, for the integral
radius of $0.1^{\circ}$, $0.3^{\circ}$, $1^{\circ}$, $3^{\circ}$, 
$10^{\circ}$, $90^{\circ}$, $180^{\circ}$ respectively around the halo center, 
assuming an exposure of $3\times10^{11}$~cm$^2$~s for ten years of 
operation of the Fermi-LAT. The corresponding numbers of the extragalactic 
background photon emission,
according to the Fermi-LAT measurements 
\cite{Ackermann:2014usa}, are 0.001, 0.01, 0.1, 1.1, 11.8, 776.6, and 
1553.2, respectively. If the center of the subhalo is located in the
inner Galaxy direction, the corresponding diffuse background could be
higher by $10-100$ times \cite{Ackermann:2012pya}. It implies 
that the detection of the $\gamma$-ray emission from such a subhalo
is challenging (and hence unconstrained) to the Fermi-LAT in light of the
small number of photons and a very long exposure time. The future
ground based Cherenkov Telescope Array (CTA; \cite{Consortium:2010bc})
may be able to detect such an extended $\gamma$-ray source and test 
our model.

The Fermi-LAT $\gamma$-ray observations of the Milky Way halo set an upper 
limit of $\sv\lesssim5\times10^{-24}$ cm$^3$~s$^{-1}$ for $m_{\psi}\sim 1.5$ 
TeV, presuming Majorana DM which annihilates into $\mu^+\mu^-$ only.
The DAMPE-favored parameter region 
is completely free from this constraint. 

\item IceCube $\nu$ data 

The IceCube observations of neutrinos from the Galactic center region 
give upper limits on the DM annihilation cross-sections (again assuming
Majorana DM) of $9.6\times10^{-23}$ cm$^3$~s$^{-1}$ and
$2.6\times10^{-22}$ cm$^3$~s$^{-1}$ for the $\mu^+\mu^-$ and $\nu\bar{\nu}$
channels, respectively \cite{Aartsen:2017ulx}. These values are 
much larger than what is required to explain the DAMPE data, and no constraints can be imposed
on our model from the Galactic center neutrinos.
On the other hand, the subhalo itself may also be visible to IceCube.
The DM annihilation rate within the halo can be characterized by $\mathcal{Q}=
\int\rho^2\,dl\,d\Omega$, where $\rho$ is the density distribution, 
$l$ is the line-of-sight path length, and $\Omega$ is the integral
solid angle. The annihilation rate of the subhalo for an opening angle cone of $10^{\circ}$
 is around two times higher 
 than that of the
Galactic center. It implies the previous bounds on the cross-sections will be improved by
a factor of 2 in the presence of the subhalo. The favored region is, however, 
far below the new bounds.
All in all, the current IceCube sensitivity is not able to constrain the parameter region yet.

\item LZ sensitivity 

As shown in the previous section, the preferred regions to account for the DAMPE bump and 
to reproduce the correct relic density are centered around $m_{Z'} \sim 2.6$ TeV with 
$g_{e \mu} \sqrt{q_\psi} \sim 0.1$.
Therefore, a large part of parameter space is unaffected by the PandaX-II search.
 The next generation DM experiment LUX-ZEPLIN (LZ)~\cite{Malling:2011va,Cushman:2013zza}, however, can
further improve the bound on the DM-nucleon cross-section by a factor of 50 or so,
 i.e., $\sigma^{\text{SI}}_{\chi p} \sim 2.4\times 10^{-11}$ pb for TeV DM, before reaching the neutrino floor.
 It implies
\begin{align}
g^2_{e \mu} q_\psi \left( \frac{3058 \, \text{GeV}}{m_{Z'}} \right)^2 \lesssim 1 \, ,
\end{align}
as indicated by the red contours in the Fig.~\ref{fig:combined}. In other words, the LZ can probe a sizable part of the preferred region.

\item ILC sensitivity 
 
The LEP measurements on $e^+ e^- \to e^+ e^- , \,
 \mu^+ \mu^-$require the 
effective scale of new physics $\Lambda$~(which contributes to these processes)
to be above $20$~TeV. Future $e^+ e^-$ colliders, 
such as ILC~\cite{Baer:2013cma}, FCC-ee~(formerly known as 
TLEP~\cite{Gomez-Ceballos:2013zzn}) and 
CEPC~\cite{CEPC-SPPCStudyGroup:2015csa}, can further improve the limit.
The ILC, for instance, with an integrated luminosity of 1000 fb$^{-1}$, 
can probe the new physics scale $\Lambda$ beyond 75 
TeV~\cite{Riemann:2001bb,Baer:2013cma} via the process $e^+ e^- \to 
\mu^+ \mu^-$, leading to the bound $m_{Z'}/g_{e\mu}\gtrsim21$ TeV. 
The precise value of the lower bound depends on systematic uncertainties 
and the polarization of the electron and positron beams at the ILC. 

As shown in Fig.~\ref{fig:combined}, the combination of ILC and LZ projected sensitivities can
disfavor a large region of the parameter space. Assuming ILC and LZ find no evidence
of new physics, only the resonance region ($2 m_\psi \approx m_{Z'}$) remains viable.

\end{itemize}

\section{Conclusion}
%{\it Conclusion ---}
In this work, we propose a simple $U(1)_{e \mu}$ 
flavored neutrino mass model inspired by the DAMPE $e^+ + e^-$ 
excess at energies around $1.4$ TeV~\cite{Ambrosi:2017wek}.
The first two generations of leptons are charged under $U(1)_{e \mu}$
while the third one is neutral. 
After $U(1)_{e \mu}$ and electroweak symmetry breaking, the right-handed 
neutrino Majorana mass matrix is featureless,
while the neutrino Dirac mass matrix is diagonal in the flavor basis.
The observed neutrino masses and mixing angles can hence be easily realized 
via the Type-I seesaw mechanism. 

The DM particle, a $U(1)_{e \mu}$-charged vector fermion $\psi$, annihilates 
into electrons, muons and neutrinos. To account for the DAMPE 
excess, a local DM subhalo with a mass of $M_{\rm sub}=1.25\times10^6$ 
M$_{\odot}$ at a distance of 0.1 kpc from the Earth is needed. 
CREs lose energy so quickly on the way towards the Earth that they mostly
have to come from a nearby area. The preferred parameter
region is centered around $(m_{\psi'}, \, m_{Z'}) \sim (1.5, \, 2.6)$~TeV 
with $\sv \sim 10^{-25}$~cm$^3$~s$^{-1}$. We have scrutinized constraints 
from indirect searches~(Fermi-LAT and IceCube), direct DM searches and LEP.
Interestingly, a significant portion of the preferred parameter space is within the reach 
of the next generation lepton colliders and DM direct detection experiments.
 
 \section*{Acknowledgments}
%{\it Acknowledgements ---} 
This work is supported in part by the National Key Research and Development
Program of China (No. 2016YFA0400200), the National Natural Science Foundation
of China (Nos. 11525313, 11722328), and the 100 Talents Program of Chinese 
Academy of Sciences. WCH is supported by Danish Council for Independent Research
Grant DFF-6108-00623. The CP3-Origins centre is partially funded by the Danish National Research Foundation, grant number DNRF90.

%\bibliography{TeV_DM}
%\bibliographystyle{h-physrev}

\end{document}